\documentclass[12pt]{article}

\usepackage{latexsym}
\usepackage{amsfonts}
\usepackage{amssymb}
\usepackage{amsmath}
\usepackage{youngtab}

\def\bydef{\equiv}

\def\R{\mathbb R}

\def\be{\begin{equation}}
\def\ee{\end{equation}}
\def\bea{\begin{eqnarray}}
\def\eea{\end{eqnarray}}
\def\beal{\begin{align*}}
\def\eeal{\end{align*}}
\def\boxbox{{\text{\tiny \yng(2)}}}
\def\boxabox{{\text{\tiny \yng(1,1)}}}
\newcommand{\ket}[1]{\left|#1\right\rangle} 
\newcommand{\bra}[1]{\left\langle#1\right|} 
\begin{document}

\title{Generalization of the Gell-Mann formula for $sl(5, \R)$ and $su(5)$
  algebras
\hspace{.25mm}\thanks{\,This work was supported in part by MNTR,
Belgrade, Project-141036.}}
\author{\bf{Igor Salom}
\hspace{.25mm}\thanks{\,e-mail address: isalom@phy.bg.ac.rs} \\
\normalsize{Institute of Physics, P.O. Box 57, 11001 Belgrade, Serbia}
\vspace{2mm} \\
\bf{Djordje \v Sija\v cki}
\hspace{.25mm}\thanks{\,e-mail address: sijacki@phy.bg.ac.rs} \\
\normalsize{Institute of Physics, P.O. Box 57, 11001 Belgrade, Serbia}
\vspace{2mm} }

\maketitle

\begin{abstract}

The so called Gell-Mann formula expresses the Lie algebra elements
in terms of the corresponding In\"on\"u-Wigner contracted ones. In
the case of $sl(n, \R)$ and $su(n)$ algebras contracted w.r.t.
$so(n)$ subalgebras, the Gell-Mann formula is generally not valid,
and applies only in the cases of some algebra representations. A
generalization of the Gell-Mann formula for $sl(5,\R)$ and $su(5)$
algebras, that is valid for all representations, is obtained in a
group manifold framework of the $SO(5)$ and/or $Spin(5)$ group.

\end{abstract}

\section{Introduction}

The Gell-Mann, or "decontraction" formula is a simple prescription
designed to determine a deformation of a Lie algebra that is
"inverse" to the In\"on\"u-Wigner contraction \cite{InonuWigner}.
This formula expresses elements of "decontracted" algebra in terms
of elements of the contracted one. Since, by a rule, various
properties of the contracted algebra are much easier to explore
(e.g. construction of representations \cite{Mackey},
decompositions of a direct product of representations
\cite{HermannBook}, etc.), this formula found its place, as a
useful and simple tool, even in some textbooks \cite{HermannBook}
and in the mathematical encyclopedia \cite{EncyclMath}.

There is a myriad of In\"on\"u-Wigner Lie algebra contraction
applications arising in various parts of Theoretical Physics. Just
to mention a few ranging from contractions from the Poincar\'e
algebra to the Galilean one, and from the Heisenberg algebras to
the Abelian ones of the same dimensions (a symmetry background of
a transition processes from relativistic and quantum mechanics to
classical mechanics) to those of contractions from (Anti-)deSitter
to the Poincar\'e algebra, and various cases involving Virasoro
and Kac-Moody algebras. A recent study of an Affine Gauge Gravity
Theory in $5D$ \cite{R13a} is heavily related to the $sl(5,\R)$
algebra contraction w.r.t. its $so(1,3)$ subalgebra, and the
representations of the relevant algebras.

The main drawback of the Gell-Mann formula is its limited
validity. There is a number of references dealing with the
question when this formula is applicable \cite{HermannBook,
Hermann, Berendt}. The formula is best studied in the case of
(pseudo) orthogonal algebras $so(m,n)$ contracted w.r.t. their
$so(m-1,n)$ and/or $so(m,n-1)$ subalgebras, i.e.\ on the group
level for $SO(m,n) \rightarrow R^{m+n-1} \wedge SO(m-1,n)$ or
$SO(m,n) \rightarrow R^{m+n-1} \wedge SO(m,n-1)$, where, loosely
speaking, the Gell-Mann formula works very well \cite{Sankara}.
Moreover, the case of (pseudo) orthogonal algebras is the only one
where this formula is valid for (almost) all representations
\cite{Weimar}. There were some attempts to generalize the
Gell-Mann formula \cite{Stovicek, Mukunda}, that resulted in a
construction of relatively complicated polynomial formulas for the
"decontracted" algebra operators, applicable to complex simple Lie
algebras $g$ with respect to decomposition $g = k + i k = k_c$.

In this work we are generally interested in Gell-Mann's formula
for the $sl(n,\R)$ algebras, that are contracted w.r.t. their
maximal compact $so(n)$ subalgebras. Note that, due to mutual
relations of the $sl(n,\R)$ and $su(n)$ algebras, one can convey
the results obtained for the $sl(n,\R)$ algebras to the
corresponding ones of the $su(n)$ algebras. There are some
subtleties in that process that will be considered below. The
Gell-Mann formula is, in this case, especially valuable for the
problem of finding all unitary irreducible representations of the
$sl(n,\R)$ algebras in the basis of the $SO(n)$ and/or $Spin(n)$
groups generated by their $so(n)$ subalgebras. Finding
representations in the basis of the maximal compact subgroup
$SO(n)$ of the $SL(n,\R)$ group, is mathematically superior, and
it suites well various physical applications in particular in
nuclear physics, gravity, physics of p-branes \cite{SijackiBranes}
etc. Moreover, this framework opens up a possibility of finding,
in a rather straightforward manner, all matrix elements of
noncompact $SL(n,\R)$ generators for all representations.
Unfortunately, the original Gell-Mann formula is, in that respect,
limited to some classes of (multiplicity free) representations
only.

As already stated, the Gell-Mann formula, except in the cases of
(pseudo) orthogonal algebras, is not generally valid by itself,
and its validity depends on the representations of the algebra as
well. Therefore, in the case of the $SL(n,\R)$ groups, i.e.\ their
$sl(n,\R)$ algebras, one is faced, in addition to the pure
algebraic features, with a problem of studying the matters that
are relevant to characterize representations as well: (i) the
group topology properties, and (ii) the non trivial multiplicity
of the $SL(n,\R)$, and $sl(n,\R)$ representations in the $SO(n)$,
and $so(n)$ basis, respectively. Both features are rather subtle
for $n \geq 3$. Note that, in the case of the $sl(n,\R)$ algebras,
due to a fact that the generalization of the Gell-Mann formula
obtained below depends on the algebra representation features, we
deviate from the standard Lie algebra deformation approach.

The $SL(n,\R)$ group can be decomposed, as any semisimple Lie group,
into the product of its maximal compact subgroup $K = SO(n)$, an Abelian group
$A$ and a nilpotent group $N$. It is well known that only $K$ is not
guaranteed to be simply-connected. There exists a universal covering group
$\overline{K} = \overline{SO}(n)$ of $K = SO(n)$, and thus also a universal
covering of $G =SL(n,\R)$: $\overline{SL}(n,\R) \simeq \overline{SO}(n)
\times A \times N$. For $n \geq 3$, $SL(n,\R)$ has double covering, defined by
$\overline{SO}(n) \simeq Spin(n)$ the double-covering of the $SO(n)$
subgroup. The universal covering group $\overline{G}$ of a given
group $G$ is a group with the same Lie algebra and with a simply-connected
group manifold. A finite dimensional covering, $\overline{SL}(n,\R)$ exists
provided one can embed $\overline{SL}(n,\R)$ into a group of finite complex
matrices that contain $Spin(n)$ as subgroup. A scan of the Cartan classical
algebras points to the $SL(n,C)$ groups as a natural candidate for the
$SL(n,\R)$ groups covering. However, there is no match of the defining
dimensionalities of the $SL(n,C)$ and $Spin(n)$ groups for $n \geq 3$,
$dim(SL(n, C)) = n < 2^{\left[\frac{n-1}{2}\right] } =
dim(Spin(n))$, except for $n = 8$. In the $n = 8$ case, one finds that the
orthogonal subgroup of the $SL(8,\R)$ and $SL(8, C)$ groups is $SO(8)$ and not
$Spin(8)$. For a detailed account of the $D=4$ case cf. \cite{YN+DjS-IJMPA}.
Thus, we conclude that there are no covering groups of the $SL(n,\R)$, $n \geq
3$ groups defined in finite-dimensional spaces. An explicit construction of all
$SL(3,R)$ irreducible representations, unitary and nonunitary
multiplicity-free spinorial \cite{DjS-JMP31}, and unitary
non-multiplicity-free \cite{SijackiSL3}, shows that they are
infinite-dimensional. The universal (double) covering groups,
$\overline{SL}(n,R)$, $n \geq 3$ of the $SL(n,R)$, $n \geq 3$ group
are groups of infinite complex matrices. All their spinorial representations
are infinite dimensional. In the reduction of this representations
w.r.t. $Spin(n)$ subgroups, one finds $Spin(n)$ representations of unbounded
spin values.

The $SU(n)$ groups are compact, with a simply connected group manifold, thus
being its own universal coverings. The $SO(n)$ subgroups are embedded into the
$SU(n)$ groups as $n$-dimensional matrices, and this embedding does not allow
nontrivial (double) covering of $SO(n)$ within $SU(n)$. As a consequence, in
the reduction of the $SU(n)$ unitary irreducible representations one finds the
tensorial $SO(n)$ representations only.

An inspection of the unitary irreducible representations of the
$\overline{SL}(n,\R)$, $n = 3,4$  groups \cite{SijackiSL3, SijackiSL4} shows
that they have, as a rule,
a nontrivial multiplicity of the $Spin(n)$, $n = 3, 4$ subgroup
representations. It is well known, already from the case of the $SU(3)$
representations in the $SO(3)$ subgroup basis, that the additional labels
required to describe this nontrivial multiplicity cannot be solely related to
the group generators themselves. An elegant solution, that provides the
required additional labels, is to work in the group manifold of the $SO(n)$
maximal compact subgroup, and to consider an action of the group both to the
right and to the left. In this way one obtains, besides the maximal compact
subgroup labels, an additional set of labels to describe the $SO(n)$ subgroup
multiplicity.

All unitary irreducible representations of the $\overline{SL}(3,\R)$ and
$\overline{SL}(4,\R)$ groups are classified, and various relevant explicit
expressions are known \cite{SijackiSL3, SijackiSL4}. It turns out
that an answer to the question
of the Gell-Mann formula generalization can be effectively read out from these
known closed form expressions of representations of noncompact generators in
both $SL(3,\R)$ and $SL(4,\R)$ cases. Such a generalization then, as a rule,
has an overall validity for all representations. We study the known
representations of the noncompact $SL(3, \R)$ and $SL(4, \R)$ generators in
the maximal compact subgroup basis, and infer the sought for expressions for
the corresponding Gell-Mann formula. On the basis of these results, we turn to
the case of the $SL(5,\R)$ generators, whose unitary irreducible
representations are not known completely. As a result, we obtain a single
closed expression that generalizes the Gell-Mann formula for the $sl(5,\R)$
algebra w.r.t. its maximal compact $so(5)$ subalgebra.

\section{In\"on\"u-Wigner contraction of $sl(n, R)$ algebras}

The $sl(n,\R)$ algebra operators, i.e.\ the $SL(n,\R)$ group
generators, can be split into two subsets: $M_{ab}$, $a,b = 1, 2,
...,n$ operators of the maximal compact subalgebra $so(n)$
(corresponding to the antisymmetric real $n\times n$ matrices,
$M_{ab} = -M_{ba}$), and the, so called, sheer operators $T_{ab}$,
$a,b = 1, 2, ...,n$ (corresponding to the symmetric traceless real
$n\times n$ matrices, $T_{ab} = T_{ba}$). The $sl(n,\R)$
commutation relations, in this basis, read: \bea [M_{ab},M_{cd}]
&=& i(\delta_{ac}M_{bd} +
\delta_{ad}M_{cb} - \delta_{bc}M_{ad} - \delta_{bd}M_{ca}), \\
{}[M_{ab},T_{cd}] &=& i(\delta_{ac}T_{bd} +
\delta_{ad}T_{cb} - \delta_{bc}T_{ad} - \delta_{bd}T_{ca}),
                                                 \label{MTcommutator}\\
{}[T_{ab},T_{cd}] &=& i(\delta_{ac}M_{db} + \delta_{ad}M_{cb} +
\delta_{bc}M_{da} + \delta_{bd}M_{ca}).         \label{TTcommutator}\eea %

The $su(n)$ algebra operators can be split likewise w.r.t. its $so(n)$
subalgebra into $M_{ab}$ and $T^{su(n)}_{ab}$, $a,b = 1, 2, ...,n$. The
$T^{su(n)}_{ab}$ and $T_{ab}$ operators are mutually related by
$T^{su(n)}_{ab}$ = i $T_{ab}$, and the $[T^{su(n)}_{ab}, T^{su(n)}_{cd}]$
differs from (\ref{TTcommutator}) by having an overall plus sign on the
right-hand side.

The In\"on\"u-Wigner contraction of $sl(n, \R)$ with respect to its
maximal compact subalgebra $so(n)$ is given by the limiting procedure: %
\be U_{ab} \bydef \lim_{\epsilon \rightarrow 0} (\epsilon T_{ab}),
                                                 \label{limit} \ee %
which leads to the following commutation relations: %
\bea [M_{ab},M_{cd}] &=& i(\delta_{ac}M_{bd} +
\delta_{ad}M_{cb} - \delta_{bc}M_{ad} - \delta_{bd}M_{ca}) \\
{}[M_{ab},U_{cd}] &=& i(\delta_{ac}U_{bd} +
\delta_{ad}U_{cb} - \delta_{bc}U_{ad} - \delta_{bd}U_{ca})
                                                  \label{MUcommutator} \\
{}[U_{ab},U_{cd}] &=& 0. \eea %

Therefore, the In\"on\"u-Wigner contraction of $sl(n, \R)$ gives a semidirect
sum  $r_{\frac{n(n+1)}{2}-1}\biguplus so(n)$ algebra, where
$r_{\frac{n(n+1)}{2}-1}$ is an Abelian subalgebra (ideal) of ``translations''
in $\frac{n(n+1)}{2}-1$ dimensions.

The Gell-Mann formula, which is a prescription to provide an "inverse" to the
contraction, (\ref{limit}), in this case reads: %
\be T_{ab} = \sigma U_{ab} + \frac{i\alpha}{\sqrt{U\cdot U}}
[C_2(so(n)), U_{ab}],                          \label{Gell-Mann_original}\ee %
where $C_2(so(n))$ denotes the second order Casimir operator of the $so(n)$
algebra, $\frac 12 \sum M_{ab} M_{ab}$, while $\sigma$ is an arbitrary
(complex) parameter and $\alpha$ is a (real) normalization constant that
depends on $n$.

In order to make use of the Gell-Mann formula to obtain $sl(n,\R)$
representations, the first necessary step is to construct representations of
the contracted algebra. Representations of the $so(n)$ generators $M_{ab}$
are well known. There are two properties that characterize representations of
the $U_{ab}$ operators: (i) The $U_{ab}$ operators transform w.r.t the
$\frac{n(n+1)}{2} -1$ dimensional representation of $so(n)$
(\ref{MUcommutator}), i.e.\ as a symmetric second order $so(n)$ tensors (in
Young diagram notation $\boxbox$), and (ii) $U_{ab}$ mutually commute.
These two requirements are met by expressing the $U_{ab}$ operators in terms
of the, so called, Wigner $D$-function (the $SO(n)$ group
matrix elements  expressed as functions of the group parameters): %
\be U_{ab} \sim D^{(\boxbox)}_{(cd)(ab)}\!(g^{-1}(\theta)) \equiv
\left<
\begin{array}{c} \boxbox \\(cd) \end{array} \right| g^{-1}(\theta)
\left| \begin{array}{c} \boxbox \\(ab) \end{array} \right> ,
\label{UisD}\ee $g(\theta)$ being an $SO(n)$ element parameterized
by $\theta$, pairs of indices $(ab)$ and $(cd)$ label the $SO(n)$
matrix elements, while $\left| \begin{array}{c} \boxbox \\(ab)
\end{array} \right>$ form a basis of the $\boxbox$ representation
space. Taking inverse of $g$ in (\ref{UisD}) insures the correct
transformation properties.

The contracted $r_{\frac{n(n+1)}{2}-1}\biguplus so(n)$ and
$sl(n,\R)$ algebras are represented in a space of square
integrable functions over the $Spin(n)$ group (in accord with the
$SL(n,\R)$ topological properties), with a standard invariant Haar
measure: ${\cal L}^2(Spin(n))$. Harish-Chandra proved
\cite{HarishChandra} that this space is rich enough to contain all
possible representations (up to equivalence) of the
$\overline{SL}(n,\R)$ group, i.e.\ $sl(n,\R)$ algebra. The
$U_{ab}$ operators act multiplicatively on this space, while the
$so(n)$ subalgebra
operators act, in a standard way, via a left group action: %
$$
M_{ab} \ket{\phi} = -i \frac{d}{dt} \exp(i t M_{ab})\Big|_{t=0}
\ket{\phi}, \quad g' \ket{g} = \ket{g'g},
\quad \ket{\phi} \in {\cal L}^2(Spin(n)).
$$
This representation space is highly reducible, however this fact is of no
relevance for the present considerations.

The $U_{ab}$ expressions (\ref{UisD}) fulfill, straightforwardly, both required
properties: commute mutually, as being ordinary functions of $\theta$, and
transform under $\boxbox$ when acting to the right on a ket vector
$\left| \begin{array}{c} \boxbox \\ (ab) \end{array} \right>$
(characterized accordingly by the $(ab)$ indices). However, the bra vector
$\left< \begin{array}{c} \boxbox \\ (cd) \end{array} \right|$, standing to the
left of $R(\theta)$, can be an arbitrary vector from the $\boxbox$
representation, thus providing an additional set of labels of the $U_{ab}$,
and accordingly the $T_{ab}$, operators. It turns out, as it will be seen
below, that these additional labels play an important role in the $sl(n,\R)$
unitary irreducible representations description, in particular in
characterizing nontrivial $Spin(n)$ subgroup multiplicity.

A natural orthonormal basis in the $Spin(n)$ representation space is given by
properly normalized functions of the $Spin(n)$ representation matrix elements:%
\be \left\{\left| { \begin{array}{l@{}l}  J & \\ k & m
\end{array}} \right> \equiv \int {\scriptstyle \sqrt{dim(J)}}
D^J_{km}\!(g(\theta)^{-1}) d\theta \ket{g(\theta)}\right\}, \
\left< { \begin{array}{l@{}l|l@{}l}  J & & J'\\ k & m & k' & m'
\end{array}} \right> = \delta_{JJ'}\delta_{kk'}\delta_{mm'},
\label{naturalbasis}\ee where $d\theta$ is an (normalized)
invariant Haar measure, and $D^{J}_{km}$
are the representation matrix elements %
$$
D^{J}_{km}(\theta) \equiv
\left< \begin{array}{c} J \\k \end{array} \right| R(\theta)
\left| \begin{array}{c} J \\m \end{array} \right>.
$$ %
Here, $J$ stands for a set of $Spin(n)$ irreducible representation labels,
while $k$ and $m$ labels numerate representation basis vectors.

An action of the $so(n)$ operators in this basis is well known, and it can be
written in terms of the Clebsch-Gordan coefficients of the $Spin(n)$ group as
follows,
{\renewcommand{\arraystretch}{0.2} %
\be \left< M_{ab}\right> = \left< { \begin{array}{l@{}l} J' & \\
k' & m' \end{array}} \right| M_{ab} \left| {\begin{array}{l@{}l} J
& \\ k & m \end{array}} \right> = \delta_{JJ'} {\scriptstyle
\sqrt{C_2(J)}} \; C\!\!\!{\scriptsize
\begin{array}{c@{}c@{}c} J & \boxabox & J' \\ m & (ab) & m' \end{array}}.
                                                       \label{Maction}\ee} %
The matrix elements of the $U_{ab}$ operators in this basis are readily found
to read: %
{\renewcommand{\arraystretch}{0.2} \bea \left< U_{ab}\right> &=&
\left< { \begin{array}{l@{}l} J' & \\ k' & m' \end{array}} \right|
D^{-1\boxbox}_{(cd)(ab)} \left| {\begin{array}{l@{}l} J & \\ k & m
\end{array}}
\right> \nonumber \\ %
&=& {\scriptstyle \sqrt{dim(J')dim(J)}} \int D_{k'm'}^{J'*}\!(\theta)
D^\boxbox_{(cd)(ab)}(\theta) D_{km}^{J}(\theta) d\theta
                                                      \label{Uaction} \\
&=& {\scriptstyle \sqrt{\frac{dim(J)}{dim(J')}}} C\!\!{\scriptsize
\begin{array}{c@{}c@{}c} J & \boxbox & J' \\ k & (cd) & k' \end{array}}
C\!\!{\scriptsize
\begin{array}{c@{}c@{}c} J & \boxbox & J' \\ m & (ab) & m' \end{array}}
\nonumber \eea} %
A closed form of the matrix elements of the whole contracted algebra
$r_{\frac{n(n+1)}{2}-1}\biguplus so(n)$ representations is thus explicitly
given in this space by (\ref{Maction}) and (\ref{Uaction}).

\section{\label{sec:OriginalGM}The Gell-Mann formula for the $sl(3,\R)$ and
$sl(4,\R)$ algebras}

Let us now, equipped with the knowledge about the contracted algebra
representation, consider the Gell-Mann formula in the cases of the $sl(3,\R)$
and $sl(4,\R)$ algebras.

In the $sl(3,\R)$ algebra case, the maximal compact subgroup of the
corresponding $\overline{SL}(3,\R)$ group is $Spin(3)$, and a basis of the
$sl(3,\R)$ representation space is given by the well known set of vectors,
\be \left\{ \left| { \begin{array}{l@{}l}  J & \\ k & m \end{array}} \right>,
J = 0, \frac 12, 1, \frac 32, \dots ; \quad |k|, |m| \leq J \right\}.
                                                   \label{sl3 basis}\ee

The traceless symmetric tensor $\boxbox$ transforms under a
five-dimensional $SO(3)$, i.e.\ $Spin(3)$, representation of
$J=2$. One can make use of an arbitrary vector from this
representation to evaluate the $U_{ab}$ operators expressions. We
take the simplest, however for our purposes adequate, realization
of the $U_{ab}$ operators, $U_{ab} \sim D^{(\boxbox)}_{(33)(ab)}$,
$a,b = 1,2,3$, i.e.\ in the spherical $SO(3)$ basis, $U_\mu \sim
D^{(\boxbox)}_{0\mu}$, $\mu = 0, \pm 1, \pm 2$. The Gell-Mann
formula
(\ref{Gell-Mann_original}) yields now:%
\be T_{\mu} = \sigma D^2_{0\mu} + i\alpha
[C_2(so(3)), D^2_{0\mu}],\quad \mu = 0,\pm 1,\pm 2\label{Gell-MannSL(3)} \ee %
The matrix elements of the shear operators $T_\mu$ are given by the following
expression: %
\be \bra{\begin{array}{l@{}l}  J' & \\ k' & m' \end{array}} T_\mu
\ket{\begin{array}{l@{}l}  J & \\ k & m \end{array}} =
{\scriptstyle\left(\vphantom{|^|}\sigma + i\alpha \left( J'(J'+1)-
J(J+1)\right) \right) \sqrt{\frac{2J+1}{2J'+1}}} C\!\!{\scriptsize
\begin{array}{c@{}c@{}c} J & 2 & J' \\ k & 0 & k'\end{array}}
C\!\!{\scriptsize \begin{array}{c@{}c@{}c} J & 2 & J' \\ m & \mu &
m'
\end{array}}                                          \label{TMatrixSL(3)}\ee

The shear operators (\ref{Gell-MannSL(3)}) satisfy the relation
(\ref{MTcommutator}) by a construction. However the relation
(\ref{TTcommutator}) is not a priory granted, and it must be
checked to hold. An explicit calculation shows that this relation
does not hold in general. It turns out, that the commutation
relations of the $sl(3,\R)$ algebra, as realized by
(\ref{Maction}) and (\ref{TMatrixSL(3)}), hold only for the
representation subspaces characterized by $k = 0$, and provided
$\alpha = 1/\sqrt{6}$, i.e.\ for \be \left\{ \left| {
\begin{array}{l@{}l}  J & \\ 0 & m \end{array}} \right>,
\quad J = 0, \frac 12, 1, \frac 32, \dots ,\quad |m| \leq J  \right\},\ee %
Therefore, the Gell-Mann formula is valid only for the $sl(3,\R)$
representations defined in the Hilbert spaces over the
$SO(3)/SO(2)$ coset space. These representations are the
multiplicity free ones w.r.t. the compact $so(3)$ subalgebra,
since the label $k$, which counts the $so(3)$ representations
multiplicity (i.e.\ in the physical terms the spin $J$ degeneracy)
is fixed.

The Gell-Mann prescription (\ref{Gell-MannSL(3)}) does not work in the general
case; a comparison with the complete classification of the $sl(3,\R)$
representations \cite{SijackiSL3} reveals that the $sl(3,\R)$ representations
with nontrivial multiplicity, as well as the spinorial multiplicity free
representations, cannot be obtained in this way ($k=0$ implies that $J$ must
take strictly integer values). Moreover, a detailed analysis shows that this
conclusion hold irrespectively of the concrete choice $U_{\mu} \sim
D^{\boxbox}_{0\mu}$ that we have made.

In the $sl(4,\R)$ algebra case, the maximal compact subgroup of the
corresponding $\overline{SL}(4,\R)$ group, is $Spin(4)$. One possibility
of choosing a basis for the $so(4)$ representations space corresponds to the
$so(4)$ algebra decomposition $so(4) = so(3) \oplus so(3)$. The $so(4)$
representation space basis is now
$\left\{ \ket{\begin{array}{l@{}l} J_1 & J_2\\ m_1 & m_2 \end{array}}
\right\}$, where pairs $(J_1, m_1)$ and $(J_2, m_2)$ specify vectors of the
two $SO(3)$ representations. A basis of the $sl(4,\R)$ representation space
(\ref{naturalbasis}) is then given by vectors
\be \left\{ \left| { \begin{array}{l@{}l}  J_1 J_2 & \\
k_1 k_2 & m_1 m_2 \end{array}} \right>,\quad  J_i = 0, \frac 12, \dots ,\quad
|k_i|, |m_i| \leq J_i,\quad  i =1,2  \right\}.          \label{sl4basis_1}\ee

Similarly, as in the $sl(3)$ case, the Gell-Mann formula yields:%
\be T_{\mu_1\mu_2} = \sigma D^{11}_{00\mu_1\mu_2} +
\frac{i}{2} [C_2(so(4)), D^{11}_{00\mu_1\mu_2}],\quad \mu = 0,\pm 1;\
i=1,2.                                        \label{Gell-MannSL(4_1)}\ee %
The corresponding matrix elements of the $T_{\mu_1\mu_2}$ operators read: %
\bea &&\bra{\begin{array}{l@{}l}  J_1'J_2'  \\ k_1'k_2' & m_1'm_2' \end{array}}
T_{\mu_1\mu_2} \ket{\begin{array}{l@{}l}  J_1 J_2 & \\
k_1 k_2 & m_1 m_2 \end{array}}  \nonumber \\
&& = {\scriptstyle\left(\vphantom{|^|} \sigma + i
({J_1'(J_1'+1)+J_2'(J_2'+1)-J_1(J_1+1)-J_2(J_2+1)})\right)
\sqrt{\frac{(2J_1+1)(2J_2+1)}{(2J_1'+1)(2J_2'+1)}}} \times \\
&&\quad\  C\!\!{\scriptsize
\begin{array}{c@{}c@{}c} J_1 & 1 & J_1'
\\ k_1 & 0 & k_1'
\end{array}}
C\!\!{\scriptsize
\begin{array}{c@{}c@{}c} J_2 & 1 & J_2'
\\ k_2 & 0 & k_2'
\end{array}}
C\!\!{\scriptsize
\begin{array}{c@{}c@{}c} J_1 & 1 & J_1'
\\ m_1 & \mu_1 & m_1'
\end{array}}
C\!\!{\scriptsize
\begin{array}{c@{}c@{}c} J_2 & 1 & J_2'
\\ m_2 & \mu_2 & m_2'
\end{array}}. \nonumber\eea %

However, analogously as in the $sl(3,\R)$ case, the commutation relations of
the noncompact operators close correctly only for the subspaces
(\ref{sl4basis_1})  characterized by $k_1 = k_2 = 0$, and thus for $J_1$,
$J_2$ being integers. The corresponding $sl(4,\R)$ representations are the
multiplicity free ones, while the representation spaces are Hilbert spaces over
the $(Spin(3)/Spin(2))\times (Spin(3)/Spin(2))$ coset space ($Spin(2)$
denoting the double cover of $SO(2)$).

Additional $sl(4,\R)$ multiplicity free representations can be obtained
by working in the $Spin(4)$ representation spaces characterized by the
subgroup chain $Spin(4)\supset Spin(3) \supset Spin(2)$, with basis
vectors:
\be \left\{ \left| { \begin{array}{c}  J_1 J_2 \\
J \\ m \end{array}} \right>,\  J_i = 0, \frac 12, \dots ;\
|J_1 -J_2| \leq J \leq J_1 + J_2;\  |m| \leq J;\  i = 1,2 \right\}.
                                                     \label{so4basis_2}\ee

The $sl(4,\R)$ representation space is then defined by basis vectors:
\be \left\{ \left| { \begin{array}{c@{}c}  J_1 J_2 & \\
K & J \\ k & m \end{array}} \right>, J_i = 0, \frac 12, \dots, |J_1 - J_2|
\leq K,J \leq J_1 + J_2, |m| \leq J \right\}.       \label{sl4basis_2}\ee

The corresponding Gell-Mann formula expression for the (noncompact) shear
generators reads %
\be T_{j\mu} = \sigma D\!\!{\tiny
\begin{array}{c@{}c} 1 1 &  \\ 0 & j \\ \vphantom{\underline{m_1}}0 & \mu
\end{array}}
 - {\scriptstyle \frac{i\sqrt{3}}{4}}
[C_2(so(4)), D\!\!{\tiny \begin{array}{c@{}c} 1 1 &  \\ 0 & j \\
\vphantom{\underline{m_1}}0 & \mu \end{array}}], \quad j=0,1,2;\  |\mu |\leq j,
                                               \label{Gell-MannSL(4_2)}\ee %
and yields the correct commutation relations in the subspace of
(\ref{sl4basis_2}) for $K=k=0$, only.
The corresponding $sl(4,\R)$ representations are the multiplicity free ones,
defined in symmetric spaces over the $Spin(4)/Spin(3)$ coset space. These
representations have only the "diagonal" $so(4)$ content, since the
condition $K=0$ implies $J_1 = J_2$.

Note that, neither the representations with multiplicity, nor the spinorial
multiplicity free $sl(4,\R)$ representations (cf. \cite{SijackiSL4}) can be
obtained by making use of the expressions given by (\ref{Gell-MannSL(4_1)}) or
(\ref{Gell-MannSL(4_2)}).

It is clear from these examples, that the Gell-Mann formula has a
limited scope when applied to the $sl(n,\R)$, $n = 3,4$ cases. It
is neither valid as an operator expression, nor it holds for a
generic $sl(n,\R)$, $n = 3,4$ representation space. One can make
use of the Gell-Mann formula in some subspaces of the most general
representation space, only. More precisely, the expressions
(\ref{Gell-MannSL(3)}), (\ref{Gell-MannSL(4_1)}) and
(\ref{Gell-MannSL(4_2)}) yield the noncompact algebra operators in
the symmetric spaces over $Spin(3)/Spin(2)$,
$Spin(4)/(Spin(2)\times Spin(2))$ and $Spin(4)/Spin(3)$,
respectively. Note, that this result is in agreement with the
Hermann theorem \cite{Hermann} stating that the Gell-Mann formula
certainly works in the symmetric spaces $K/L$ if $K$ is a simple
compact subgroup from a Cartan decomposition of the starting group
$G$ (here ${G} = SL(n,\R)$) and if there exists some $U_{\mu}$
which is invariant under the action of $L$. Unfortunately, this
theorem does not give the necessary conditions for the Gell-Mann
formula to hold.

It turns out that we can shed some light on the question of a validity of the
Gell-Mann formula in the $sl(n,\R)$, $n=3,4$ algebra case. In doing that, we
recall first some relevant facts about the $SO(n)$, $n=3,4$ group manifold and
its representations. Let us consider an action of the $SO(n)$ representation
on the lower left-hand side labels (quantum numbers) of the basis vectors
given by (\ref{naturalbasis}). First, let us make use of the operator
\be K_{\mu}
\equiv g^{\nu\lambda} D_{\mu\nu}^\boxabox M_\lambda, \quad \mu = 1,2,\dots,n
                                                  \label{Koperator}\ee %
where $g^{\nu\lambda}$ is the Cartan metric tensor of $SO(n)$). The $K_{\mu}$
operators behave exactly as the rotation generators $M_{\mu}$, it is only
that they act on the lower left-hand side indices of the basis
(\ref{naturalbasis}):
{\renewcommand{\arraystretch}{0.2} %
\be \left< K_{ab}\right> =
\left< { \begin{array}{l@{}l} J' & \\ k' & m' \end{array}} \right| K_{ab}
\left| {\begin{array}{l@{}l} J & \\ k & m \end{array}} \right>
= \delta_{JJ'} {\scriptstyle \sqrt{C_2(J)}} \; C\!\!\!{\scriptsize
\begin{array}{c@{}c@{}c} J & \boxabox & J' \\ k & (ab) & k' \end{array}}.
                                                      \label{Kaction}\ee} %
Due to the fact that mutually contragradient $SO(n)$
representations are equivalent, the $K_{\mu}$ operators are
directly related to the "left" action of the $SO(n)$ subgroup on
${\cal L}^2(\ket{g(\theta)})$: $g' \ket{g} = \ket{g {g'}^{-1}}$.
The operators $K_{\mu}$ and $M_{\mu}$ mutually commute, however,
the corresponding Casimir operators match, i.e.\ $K_{\mu}^2 =
M_{\mu}^2$. Whereas the $sl(n,\R)$ operators $M_{ab}$ are
invariant under this left action, the shear operators $T_{ab}$,
constructed by using the Gell-Mann formula prescription, are not.
The transformation properties of the shear operators $T_{ab}$ are
inherited from the corresponding contracted operators $U_{ab}$,
which have additional nontrivial transformation properties
described by the $K_{\mu}$ operator labels (of the right-hand side
vector of the $SO(n)$ matrix in (\ref{UisD})). Consequently, a
commutator of two such shear operators has a nontrivial properties
under the $SO(n)_{K}$ group (generated by the $K_{\mu}$
operators). It is not an $SO(n)_{K}$ scalar (unlike $M_{\mu}$),
and therefore not bound to close upon the $sl(n,\R)$ commutation
relations. This is precisely the reason why the shear generators
(\ref{Gell-MannSL(3)}), (\ref{Gell-MannSL(4_1)}) and
(\ref{Gell-MannSL(4_2)}) do not satisfy the commutation relations
(\ref{TTcommutator}) in a generic representation space over the
$SO(n)$ group manifold. In particular cases, it is possible to
make a restriction to such subspace of the representation space,
where only the invariant part of the $[T, T]$ commutator survives
that is proportional to the $so(n)$ operators representation in
that subspace. For example, in the $sl(3,\R)$ case above, a
restriction was made to the subspace of $k=0$, where  the $[T, T]$
commutator piece proportional to $K_0$ vanishes; likewise for the
commutators of the shear generators (\ref{Gell-MannSL(4_1)}) and
(\ref{Gell-MannSL(4_2)}).

\section{\label{sec:GGMsl3sl4}Generalization of the Gell-Mann formula for
$sl(3,\R)$ and $sl(4,\R)$ algebras}

The above analysis raises the question of whether it is possible to modify
the Gell-Mann formula by adding some terms proportional to the generators of
the left $SO(n)_{K}$ group, that cancel the unwanted terms and "fix" the
$sl(n,\R)$ algebra commutation relations.

Such a generalization of the Gell-Mann formula in the $sl(3,\R)$ case can be
read out directly from the known matrix elements of the $sl(3,\R)$
representations with multiplicity \cite{SijackiSL3}: %
\be T_{\mu} = \sigma D^2_{0\mu} + \frac{i}{\sqrt{6}} [C_2(so(3)),
D^2_{0\mu}] + i(D^2_{2\mu}-D^2_{-2\mu})K_0 + \delta
(D^2_{2\mu}+D^2_{-2\mu}),
                                      \label{generalized sl3 GM formula}\ee %
$\mu = 0, \pm 1, \pm 2$, and where $\sigma$ and $\delta$ are the $sl(3,\R)$
group representation labels. The additional terms to the ``original''
Gell-Mann formula secure that the $T_{\mu}$ operators satisfy the commutation
relation (\ref{TTcommutator}) in the entire representation space. Note that,
there are two $sl(3,\R)$ representation labels $\sigma$ and $\delta$, matching
the algebra rank, contrary to the case of the original Gell-Mann formula whose
single free parameter cannot account for the entire representation
labeling. The additional label $k$ ($K_{0} \to k$, $|k| \leq J$) describes the
nontrivial multiplicity in $J$.

The generalized expression (\ref{generalized sl3 GM formula})
contains the original formula (\ref{Gell-MannSL(3)}) as a special
case: by restricting the representation space to the subspace of
$k = 0$, and choosing $\delta = 0$ one arrives at the multiplicity
free representations that were obtained by using the expression
(\ref{Gell-MannSL(3)}). Moreover, the generalized Gell-Mann
formula allows one to obtain some $sl(3,\R)$ multiplicity free
representations that cannot be reached by making use of the
original formula (\ref{Gell-MannSL(3)}). For example, the choice
$\sigma = \frac 32$, and $\delta = -\frac 12$ \cite{R13b} in a new
basis of vectors (linear combinations of basis vectors with
different $k$ values), \be \tiny
\begin{array}{c} \Bigg\{| \begin{array}{c} \frac12 \\ m \end{array}
  >' = | { \begin{array}{c@{\ }c}  \frac12 & \\
\frac12 & m \end{array}} > + | { \begin{array}{c@{\ }c} \frac12 &
\\
-\frac12 & m \end{array}} >, \\ %
| \begin{array}{c} \frac52 \\ m \end{array} >' = | {
    \begin{array}{c@{\ }c}  \frac52 & \\
\frac52 & m \end{array}} > + {\textstyle \sqrt{\frac 52}}| {
\begin{array}{c@{\ }c}  \frac52 & \\
\frac12 & m \end{array}} > + {\textstyle \sqrt{\frac 52}}| {
\begin{array}{c@{\ }c}  \frac52 & \\
-\frac12 & m \end{array}} > + | { \begin{array}{c@{\ }c} \frac52 &
\\
-\frac52 & m \end{array}} >, \nonumber \\ %
| \begin{array}{c} \frac92 \\ m \end{array} >' = | {
    \begin{array}{c@{\ }c}  \frac92 & \\
\frac92 & m \end{array}} > + | { \begin{array}{c@{\ }c} \frac92 &
  \\
\frac52 & m \end{array}} > + {\textstyle \sqrt{\frac{7}{2}}}| {
\begin{array}{c@{\ }c}  \frac92 & \\
\frac12 & m \end{array}} > + {\textstyle \sqrt{\frac{7}{2}}}| {
\begin{array}{c@{\ }c}  \frac92 & \\
-\frac12 & m \end{array}} > + | { \begin{array}{c@{\ }c} \frac92 &
  \\
-\frac52 & m \end{array}} > + | { \begin{array}{c@{\ }c} \frac92 &
  \\
-\frac 92 & m \end{array}} >, \dots \Bigg\}. \end{array} \ee %
yields a representation space without multiplicity that is closed
under the action of the $T_{\mu}$ operators. This is a basis of a
spinorial $sl(3,\R)$ unitary irreducible ($J$ content are half-odd
integers) representation space, where the original Gell-Mann
formula does not apply.

The Gell-Mann formula can similarly be generalized in the case of the
$sl(4,\R)$ algebra. Again, by extracting from the known matrix elements of the
$sl(4,\R)$ representations with multiplicity \cite{SijackiSL4}, we find: %
\bea &&T_{\mu_1\mu_2} = i\Big(\sigma D^{11}_{00\mu_1\mu_2} +
{\textstyle \frac{1}{2}}[C_2(so(4)), D^{11}_{00\mu_1\mu_2}] \nonumber \\
&& + \delta_1 (D^{11}_{11\mu_1\mu_2} + D^{11}_{-1-1\mu_1\mu_2})
+(D^{11}_{11\mu_1\mu_2} -
D^{11}_{-1-1\mu_1\mu_2})(K^{10}_{00}+K^{01}_{00})  \nonumber \\
&& + \delta_2 (D^{11}_{-11\mu_1\mu_2} + D^{11}_{1-1\mu_1\mu_2})
+(D^{11}_{-11\mu_1\mu_2} -
D^{11}_{1-1\mu_1\mu_2})(K^{10}_{00}-K^{01}_{00})\Big),
                                   \label{generalized Gell-MannSL(4_1)} \eea %
where $\mu_1 , \mu_2 = 0, \pm 1$. As the rank of the $sl(4, \R)$ algebra is
three, there are precisely three  representation labels $\sigma$, $\delta_1$,
and $\delta_2$ (if complex, only three real are independent).

As in the $sl(3, \R)$ case, the generalized formula reduces, for certain values
of the labels, in a representation subspace defined by $k_1 = k_2 = 0$  to the
original Gell-Mann formula (\ref{Gell-MannSL(4_1)}). It is not a
straightforward matter to see that the formula (\ref{Gell-MannSL(4_2)})
follows from the generalized Gell-Mann formula. However, the generalized
Gell-Mann formula for the $sl(4,\R)$ algebra can be expressed in an equivalent
form as follows: %
\bea && T_{j\mu} = \gamma_1 D\!\!{\tiny
\begin{array}{c@{}c}
1 1 &  \\ 0 & j \\
\vphantom{\underline{m_1}}0 & \mu
\end{array}}
 - {\scriptstyle \frac{i\sqrt{3}}{4}}
[C_2(so(4)), D\!\!{\tiny
\begin{array}{c@{}c}
1 1 &  \\ 0 & j \\
\vphantom{\underline{m_1}}0 & \mu
\end{array}}]   \nonumber \\ %
&& + \gamma_2 D\!\!{\tiny
\begin{array}{c@{}c}
1 1 &  \\ 2 & j \\
\vphantom{\underline{m_1}}0 & \mu
\end{array}} +
{\scriptstyle i\sqrt{2}} D\!\!{\tiny
\begin{array}{c@{}c}
1 1 &  \\ 2 & j \\
\vphantom{\underline{m_1}}1 & \mu
\end{array}}( K\!\!{\tiny
\begin{array}{c}
1 0  \\ 1 \\
\vphantom{\underline{m_1}} -1
\end{array}}+K\!\!{\tiny
\begin{array}{c}
0 1  \\ 1 \\
\vphantom{\underline{m_1}} -1
\end{array}})  -
{\scriptstyle i\sqrt{2}} D\!\!{\tiny
\begin{array}{c@{}c}
1 1 &  \\ 2 & j \\
\vphantom{\underline{m_1}}-1 & \mu
\end{array}}( K\!\!{\tiny
\begin{array}{c}
1 0  \\ 1 \\
\vphantom{\underline{m_1}} 1
\end{array}}+K\!\!{\tiny
\begin{array}{c}
0 1  \\ 1 \\
\vphantom{\underline{m_1}} 1
\end{array}})  \label{generalized Gell-MannSL(4_2)} \\ %
&& + \gamma_3 (D\!\!{\tiny
\begin{array}{c@{}c}
1 1 &  \\ 2 & j \\
\vphantom{\underline{m_1}}2 & \mu
\end{array}} + D\!\!{\tiny
\begin{array}{c@{}c}
1 1 &  \\ 2 & j \\
\vphantom{\underline{m_1}}-2 & \mu
\end{array}}) +
i(D\!\!{\tiny
\begin{array}{c@{}c}
1 1 &  \\ 2 & j \\
\vphantom{\underline{m_1}}2 & \mu
\end{array}} - D\!\!{\tiny
\begin{array}{c@{}c}
1 1 &  \\ 2 & j \\
\vphantom{\underline{m_1}} -2 & \mu
\end{array}}) (K\!\!{\tiny
\begin{array}{c}
1 0  \\ 1 \\
\vphantom{\underline{m_1}} 0
\end{array}}+K\!\!{\tiny
\begin{array}{c}
0 1  \\ 1 \\
\vphantom{\underline{m_1}} 0
\end{array}}),  \nonumber \eea %
where $j = 0,1,2$, $|\mu | \leq j$, and provided $i\sigma =
-\frac{1}{\sqrt{3}}\gamma_1 + \sqrt{\frac 23}\gamma_2 - 2i$,
$\delta_1 = \gamma_3$, and $\delta_2 = \frac{1}{\sqrt{3}}\gamma_1
+ \frac{1}{\sqrt{6}}\gamma_2 -2i$. Derivation of
(\ref{Gell-MannSL(4_2)}) is now obvious for $\gamma_2 = \gamma_3 =
0$. In a parallel to the $sl(3, \R)$ algebra case, the generalized
Gell-Mann formula for the $sl(4,\R)$ algebra case holds for all
representations, irrespectively of the $su(4)$ representations
multiplicity.

It is now straightforward to write down the generalized Gell-Mann formula
expressions for the $su(3)$ and $su(4)$ algebras, thus obtaining the
generators of the $SU(3)/SO(3)$ and $SU(4)/SO(4)$ factor groups.

$su(3)$: $T^{su(3)}_{\mu}$ = $i T_{\mu}$, $\mu = 0,\pm 1, \pm 2$, where
  $T_{\mu}$ is given by (\ref{generalized sl3 GM formula}).

$su(4)$: $T^{su(4)}_{\mu_1\mu_2}$ = $i T_{\mu_1\mu_2}$, $\mu_1 ,\mu_2 = 0,
 \pm 1$, i.e.\ $T^{su(4)}_{j\mu} = i T_{j\mu}$, $j =0,1,2$, $\mu \leq |j|$,
where $T_{\mu_1 \mu_2}$ and $T_{j\mu}$ are given by (\ref{generalized
 Gell-MannSL(4_1)}) and (\ref{generalized Gell-MannSL(4_2)}), respectively.

\section{\label{sec:GGMsl5}Generalized Gell-Mann formula in the $sl(5,\R)$
case}

We have shown above that the original Gell-Mann formula
expressions for the $sl(n,\R)$ and $su(n)$ algebras do not satisfy
the $[T, T]$ commutation relations (\ref{TTcommutator}) on a pure
algebraic level. However, by setting the Gell-Mann formula
existence question into a group representation framework, it is
possible to restrict representation spaces and thus achieve a
closure of the $[T, T]$ commutator ($T$ being given by the
Gell-Mann formula expression). Moreover, we have shown, by
extracting information from the known results about the
$sl(n,\R)$, $n = 3,4$ representations, that there exist a
generalization of the Gell-Mann formula for $sl(n,\R)$, $n = 3,4$,
which is valid for all representation spaces. An important role,
in that process, was played by the $K$ operator (\ref{Koperator}).

In the following, we make use of the Gell-Mann formula generalization for
$sl(n,\R)$, $n = 3,4$, and a peculiarity of the $so(5)$ algebra representation
labels to follow the $so(4)$ $=$ $so(3)\oplus so(3)$ labeling features.

Let us recall first some basic $so(5)$ algebra representation
notions. The $so(5)$ algebra is of rang two, and its irreducible
representations are labeled by a pair of labels $(\overline{J}_1,
\overline{J}_2 )$, resembling the $so(4)$ labeling. The complete
labeling of the representation space vectors can be achieved by
making use of the subalgebra chain: $so(5)$ $\supset$ $so(4)$ $=$
$so(3)\oplus so(3)$ $\supset$ $so(2)\oplus so(2)$. The basis of
the $so(5)$ algebra representation space can be taken as in
\cite{Hecht, SO5CGC}: %
\be \left\{ \left| { \begin{array}{cc}  \overline J_1 & \overline J_2 \\
J_1 & J_2 \\ m_1 & m_2 \end{array}} \right>, \quad \overline J_i = 0,
\frac 12, \dots; \quad \overline J_1 \geq \overline J_2; \quad |m_i| \leq J_i,
\quad i=1,2 \right\}.                                   \label{so5basis}\ee %
The admissible values of $J_1$ and $J_2$, within an irreducible
representation $(\overline J_1, \overline J_2)$ are given in
\cite{Kemmer}. Now, the basis of the $so(5)$ algebra, i.e.\ the
$Spin(5)$ group, representation space vectors
(\ref{naturalbasis}) is given as follows: %
\be \left\{ \left| { \begin{array}{cccc}  \overline J_1 & \overline J_2 & & \\
K_1 & K_2 & J_1 & J_2 \\ k_1 & k_2 & m_1 & m_2 \end{array}}
\right>\right\}.\label{sl5basis}\ee %

The ten $so(5)$ algebra operators, generating the adjoint representation of
$Spin(5)$, transform, in notation (\ref{so5basis}), under the representation
$(\overline 1,\overline 0)$. Their $so(4)$ subalgebra representation content
is: $(\overline 1,\overline 0)$ $\rightarrow$ $(1,0)$ $\oplus$ $(\frac
12,\frac 12)$ $\oplus$ $(0,1)$. The shear operators transform under the
$14$-dimensional $so(5)$ irreducible representation $(\overline 1,
\overline 1)$ of  $so(5)$ which contains $(1,1)$, $(\frac 12, \frac 12)$ and
$(0,0)$ representation upon reduction to $so(4)$: %
{\renewcommand{\arraystretch}{0.2}
$$\left\{ T \!\!{\tiny \begin{array}{l@{\!}l} \vphantom {\overline 1}
& \\ j_1 & j_2 \\
\vphantom{\underline{\mu_1}}\mu_1 & \mu_2
\end{array}}\right\} = \left\{T\!\!{\tiny \begin{array}{l@{\!}l}
\vphantom {\overline 1} &
\\ 1 & 1 \\
\vphantom{\underline{\mu_1}}\mu_1 & \mu_2
\end{array}},T\!\!{\tiny \begin{array}{l@{\!}l}
\vphantom {\overline 1} &
\\ \frac 12 & \frac 12 \\
\vphantom{\underline{\mu_1}}\mu_1 & \mu_2
\end{array}}, T\!\!{\tiny \begin{array}{l@{\!}l}
\vphantom {\overline 1} &
\\ 0 & 0 \\
\vphantom{\underline{\mu_1}}0 & 0
\end{array}}\right\}.$$}

The original Gell-Mann formula is not applicable, again, in the
whole space spanned by (\ref{sl5basis}), but only in the symmetric
spaces $Spin(5)/Spin(4)$ and $Spin(5)/(Spin(3)\otimes Spin(2))$,
with appropriate choices of the $SO(5)_K$ labels for the
contracted operators $U$. Neither representations with
multiplicity, nor spinorial representations of the $sl(5,\R)$
algebra can be obtained in this way.

We made an educated guess, based on the structure of the generalized Gell-Mann
formula for the $sl(4,\R)$ algebra, when adding possible terms to the
generalized Gell-Mann formula in the $sl(5,\R)$ case. We omit the terms
proportional to
$D\!\!{\tiny
\begin{array}{l@{}l@{}l@{}l} \overline 1 &
\overline 1 & & \\ \frac12 & \frac12 & j_1 & j_2 \\
\vphantom{\underline{m_1}}k_1 & k_2 &\mu_1 & \mu_2
\end{array}}$
altogether, as well as the terms proportional to
$D\!\!{\tiny
\begin{array}{l@{}l@{}l@{}l} \overline 1 &
\overline 1 & & \\ 1 & 1 & j_1 & j_2 \\
\vphantom{\underline{m_1}}0 & 1 &\mu_1 & \mu_2
\end{array}}$ ,
$D\!\!{\tiny
\begin{array}{l@{}l@{}l@{}l} \overline 1 &
\overline 1 & & \\ 1 & 1 & j_1 & j_2 \\
\vphantom{\underline{m_1}} 1 & 0 &\mu_1 & \mu_2
\end{array}}$ ,
$D\!\!{\tiny
\begin{array}{l@{}l@{}l@{}l} \overline 1 &
\overline 1 & & \\ 1 & 1 & j_1 & j_2 \\
\vphantom{\underline{m_1}} 0 & -1 &\mu_1 & \mu_2
\end{array}}$ , and
$D\!\!{\tiny
\begin{array}{l@{}l@{}l@{}l} \overline 1 &
\overline 1 & & \\ 1 & 1 & j_1 & j_2 \\
\vphantom{\underline{m_1}} -1 & 0 &\mu_1 & \mu_2
\end{array}}$.

The $[T, T] \subset M$ commutation relations condition
(\ref{TTcommutator}), together with a knowledge of the $so(5)$
Clebsch-Gordan coefficients for $(\overline{J_1},\overline{J_2}) =
(1,1)$ \cite{SO5CGC}, finally yields the sought for generalized
Gell-Mann formula expression for the $sl(5,\R)$ algebra
shear operators: %
{\renewcommand{\arraystretch}{0.2} %
\be\begin{array}{rl} %
T\!\!{\tiny
\begin{array}{l@{\!}l} \vphantom {\overline 1} &
\\ j_1 & j_2 \\
\vphantom{\underline{\mu_1}}\mu_1 & \mu_2
\end{array}} =&
\sigma_1 D\!\!{\tiny
\begin{array}{l@{}l@{}l@{}l} \overline 1 &
\overline 1 & & \\ 0 & 0 & j_1 & j_2 \\
\vphantom{\underline{m_1}}0 & 0 &\mu_1 & \mu_2
\end{array}} +
i\sqrt{\frac 15}[C_2(so(5)), D\!\!{\tiny
\begin{array}{l@{}l@{}l@{}l} \overline 1 &
\overline 1 & & \\ 0 & 0 & j_1 & j_2 \\
\vphantom{\underline{m_1}}0 & 0 &\mu_1 & \mu_2
\end{array}}]
\\ &
 + i\Bigg(\sigma_2 D\!\!{\tiny
\begin{array}{l@{}l@{}l@{}l} \overline 1 &
\overline 1 & & \\ 1 & 1 & j_1 & j_2 \\
\vphantom{\underline{m_1}}0 & 0 &\mu_1 & \mu_2
\end{array}} +
{\textstyle \frac 12}[C_2(so(4)_K), D\!\!{\tiny
\begin{array}{l@{}l@{}l@{}l} \overline 1 &
\overline 1 & & \\ 1 & 1 & j_1 & j_2 \\
\vphantom{\underline{m_1}}0 & 0 &\mu_1 & \mu_2
\end{array}}]
\\ &
 - D\!\!{\tiny
\begin{array}{c@{}c@{}l@{}l} \overline 1 &
\overline 1 & & \\ 1 & 1 & j_1 & j_2 \\
\vphantom{\underline{m_1}}1 & -1 &\mu_1 & \mu_2
\end{array}} (\delta_1 + K\!\!{\tiny
\begin{array}{l@{}l} \overline 1 &
\overline 0 \\ 1 & 0 \\
\vphantom{\underline{m_1}}0 & 0
\end{array}} - K\!\!{\tiny
\begin{array}{l@{}l} \overline 1 &
\overline 0 \\ 0 & 1 \\
\vphantom{\underline{m_1}}0 & 0
\end{array}})-
D\!\!{\tiny
\begin{array}{c@{}c@{}l@{}l} \overline 1 &
\overline 1 & & \\ 1 & 1 & j_1 & j_2 \\
\vphantom{\underline{m_1}}-1 & 1 &\mu_1 & \mu_2
\end{array}} (\delta_1 - K\!\!{\tiny
\begin{array}{l@{}l} \overline 1 &
\overline 0 \\ 1 & 0 \\
\vphantom{\underline{m_1}}0 & 0
\end{array}} + K\!\!{\tiny
\begin{array}{l@{}l} \overline 1 &
\overline 0 \\ 0 & 1 \\
\vphantom{\underline{m_1}}0 & 0
\end{array}})%
\\ &
 + D\!\!{\tiny
\begin{array}{c@{}c@{}l@{}l} \overline 1 &
\overline 1 & & \\ 1 & 1 & j_1 & j_2 \\
\vphantom{\underline{m_1}}1 & 1 &\mu_1 & \mu_2
\end{array}} (\delta_2 + K\!\!{\tiny
\begin{array}{l@{}l} \overline 1 &
\overline 0 \\ 1 & 0 \\
\vphantom{\underline{m_1}}0 & 0
\end{array}} + K\!\!{\tiny
\begin{array}{l@{}l} \overline 1 &
\overline 0 \\ 0 & 1 \\
\vphantom{\underline{m_1}}0 & 0
\end{array}})+
D\!\!{\tiny
\begin{array}{c@{}c@{}l@{}l} \overline 1 &
\overline 1 & & \\ 1 & 1 & j_1 & j_2 \\
\vphantom{\underline{m_1}}-1 & -1 &\mu_1 & \mu_2
\end{array}} (\delta_2 - K\!\!{\tiny
\begin{array}{l@{}l} \overline 1 &
\overline 0 \\ 1 & 0 \\
\vphantom{\underline{m_1}}0 & 0
\end{array}} - K\!\!{\tiny
\begin{array}{l@{}l} \overline 1 &
\overline 0 \\ 0 & 1 \\
\vphantom{\underline{m_1}}0 & 0
\end{array}})\Bigg)
\end{array} ,                                  \label{sl5GGM}\ee
} %
where $j_i = 0, \frac12 , 1$, $|\mu_i | \leq j_i$, $i = 1,2$, the
representation labels $\sigma_1, \sigma_2, \delta_1$ and $\delta_2$ are
arbitrary (complex) parameters (four real are independent), and $C_2(so(4)_K)$
denotes the quadratic Casimir operator of the left action $so(4)_K$ algebra.
Naturally, the same result (\ref{sl5GGM}) is obtained, when searching for the
generalized Gell-Mann formula expression, by starting with all possible
additional terms proportional to
$D\!\!{\tiny
\begin{array}{l@{}l@{}l@{}l} \overline 1 &
\overline 1 & & \\ K_1 & K_2 & j_1 & j_2 \\
\vphantom{\underline{m_1}}k_1 & k_2 &\mu_1 & \mu_2
\end{array}}$-functions
and demanding (\ref{TTcommutator}), though by a much more tedious calculation.

The $su(5)$ algebra elements, as given by the generalized Gell-Mann formula,
are the $M$ operators (\ref{Maction}), generating the $SO(5)$ subgroup of the
$SU(5)$ group, and the $iT$ operators (\ref{sl5GGM}), generating the
$SU(5)/SO(5)$ factor group.

Contrary to the generalized Gell-Mann formula for the $sl(n,\R)$,
$n=3,4$ algebras, where we started from the known matrix elements
of the shear operators, here in the case of the $sl(5,\R)$
algebra, we are in a position to obtain, for the first time, the
shear operators matrix elements for a generic representation space
starting from the generalized Gell-Mann formula expression
(\ref{sl5GGM}).

The matrix elements of the $sl(5,\R)$ shear (noncompact) operators read: %
{\renewcommand{\arraystretch}{0.2} \be \begin{array}{c}
\left< {\scriptsize \begin{array}{l@{}l@{}l@{}l} \overline J_1' & \overline
      J_2' & & \\
K_1' & K_2' & J_1' & J_2' \\ k_1' & k_2' & m_1' & m_2'
\end{array}} \right| T\!\!{\tiny \begin{array}{l@{\!}l}
\vphantom {\overline 1} &
\\ j_1 & j_2 \\
\vphantom{\underline{\mu_1}}\mu_1 & \mu_2 \end{array}}
\left| {\scriptsize \begin{array}{l@{}l@{}l@{}l} \overline J_1 & \overline J_2
      \vphantom{\overline J'_2}& & \\
K_1 & K_2 & J_1 & J_2 \vphantom{J'_2}\\ k_1 & k_2 & m_1 &
m_2\vphantom{k'_2}
\end{array}} \right> =
{\scriptstyle \sqrt{\frac{dim(\overline J_1, \overline
J_2)}{dim(\overline J'_1, \overline J'_2)}}} C\!\!{\tiny
\begin{array}{l@{\!}l@{\ }l@{\!}l@{\ }l@{\!}l} \overline J_1 & \overline J_2 &
  \overline 1 & \overline
1 & \overline J_1' & \overline J_2'
\\ J_1 & J_2 & j_1 & j_2 & J_1' & J_2' \\
\vphantom{\underline{m_1}}m_1 & m_2 &\mu_1 & \mu_2 &m_1' & m_2'
\end{array}}  \\ %
\times \Bigg( {\scriptstyle \left(\sigma_1 + i\sqrt{\frac
45}(\overline J'_1(\overline J'_1 + 2) + \overline J'_2(\overline
J'_2 + 1) - \overline J_1(\overline J_1 + 2) - \overline
J_2(\overline J_2 + 1))\vphantom{|^|}\right)}
 C\!\!{\tiny
\begin{array}{l@{\!}l@{\ }l@{}l@{\ }l@{\!}l} \overline J_1 &
\overline J_2 & \overline 1 & \overline 1 & \overline J_1' &
\overline J_2'
\\ K_1 & K_2 & 0 & 0 & K_1' & K_2' \\
\vphantom{\underline{m_1}}k_1 & k_2 & 0 & 0 &k_1' & k_2'
\end{array}}  \\
+\ {\scriptstyle i\left(\sigma_2 + K'_1(K'_1 + 1) + K'_2(K'_2 + 1)
- K_1(K_1 + 1) - K_2(K_2 + 1)\vphantom{|^|}\right)}
 C\!\!{\tiny
\begin{array}{l@{\!}l@{\ }l@{}l@{\ }l@{\!}l} \overline J_1 &
\overline J_2 & \overline 1 & \overline 1 & \overline J_1' &
\overline J_2'
\\ K_1 & K_2 & 1 & 1 & K_1' & K_2' \\
\vphantom{\underline{m_1}}k_1 & k_2 & 0 & 0 &k_1' & k_2'
\end{array}}  \\
-\ {\scriptstyle i\left(\delta_1 + k_1 - k_2\right)}
 C\!\!{\tiny
\begin{array}{l@{\!}l@{\ }l@{}l@{\ }l@{\!}l} \overline J_1 &
\overline J_2 & \overline 1 & \overline 1 & \overline J_1' &
\overline J_2'
\\ K_1 & K_2 & 1 & 1 & K_1' & K_2' \\
\vphantom{\underline{m_1}}k_1 & k_2 & 1 & -1 &k_1' & k_2'
\end{array}}
- {\scriptstyle i\left(\delta_1 - k_1 + k_2\right)}
 C\!\!{\tiny
\begin{array}{l@{\!}l@{\ }l@{}l@{\ }l@{\!}l} \overline J_1 &
\overline J_2 & \overline 1 & \overline 1 & \overline J_1' &
\overline J_2'
\\ K_1 & K_2 & 1 & 1 & K_1' & K_2' \\
\vphantom{\underline{m_1}}k_1 & k_2 & -1 & 1 &k_1' & k_2'
\end{array}}  \\
+\ {\scriptstyle i\left(\delta_2 + k_1 + k_2\right)}
 C\!\!{\tiny
\begin{array}{l@{\!}l@{\ }l@{}l@{\ }l@{\!}l} \overline J_1 &
\overline J_2 & \overline 1 & \overline 1 & \overline J_1' &
\overline J_2'
\\ K_1 & K_2 & 1 & 1 & K_1' & K_2' \\
\vphantom{\underline{m_1}}k_1 & k_2 & 1 & 1 &k_1' & k_2'
\end{array}}
+ {\scriptstyle i\left(\delta_2 - k_1 - k_2\right)}
 C\!\!{\tiny
\begin{array}{l@{\!}l@{\ }l@{}l@{\ }l@{\!}l} \overline J_1 &
\overline J_2 & \overline 1 & \overline 1 & \overline J_1' &
\overline J_2'
\\ K_1 & K_2 & 1 & 1 & K_1' & K_2' \\
\vphantom{\underline{m_1}}k_1 & k_2 & -1 & -1 &k_1' & k_2'
\end{array}}\Bigg),
\end{array}\label{TMatrixGGLSL(5)}\ee
}%
where $dim(\overline J_1, \overline J_2) = (2 \overline J_1 - 2
\overline J_2 + 1)(2 \overline J_1 + 2 \overline J_2 + 3)(2
\overline J_1 + 2)(2 \overline J_2 + 1)/6$ is the dimension of the
$so(5)$ irreducible representation characterized by  $(\overline
J_1, \overline J_2)$ \cite{Kemmer}.

To sum up, the matrix elements of the (noncompact) shear operators
$T$ (\ref{TMatrixGGLSL(5)}), together with the known matrix
elements of the (compact) $so(n)$ operators $M$ (\ref{Maction}),
define an action of the $sl(5,\R)$ algebra on the basis vectors
(\ref{sl5basis}) of representation spaces of the maximal compact
subgroup $Spin(5)$ of the $\overline{SL}(5,\R)$ group. This result
is general due to a Corollary of Harish-Chandra
\cite{HarishChandra} that explicitly applies to the case of the
$sl(5,\R)$ algebra.

\section{Conclusion}

The Gell-Mann formula, beyond the case of (pseudo) orthogonal
algebras, is not valid in general as a pure algebraic expression.
Its applicability can be broaden, by utilizing it in certain
cases, provided some Lie algebra representation conditions are
met. As for the $sl(n,\R)$ algebras, contracted w.r.t. their
$so(n)$ subalgebras, the algebraic expression of the Gel-Mann
formula matters generally for the multiplicity free
representations only. It was demonstrated in this work that one
can generalize the Gell-Mann formula for the $sl(n,\R)$, $n=3,4,5$
algebras to be valid for a generic representation space. A brief
account required by a description of the $sl(n,\R)$, $n=3,4,5$
representation spaces, that heavily depends on the corresponding
group topology properties and the $so(n)$ subgroup multiplicity,
is given. In the $sl(3, \R)$ and $sl(4, \R)$ cases we inferred,
starting from a suitable existing expressions of the algebra
operators representations with non-trivial multiplicity, a generic
generalized Gell-Mann formulas. These formulas offer a new
starting point for mathematical physics investigations, as well as
for various physical applications. By analyzing the structure of
the Gell-Mann formula for the $sl(n,\R)$, $n=3,4,5$ cases, and
making use of the specific features of the $so(5)$ algebra
representations, we obtained the generalized Gell-Mann formula for
the $sl(5,\R)$ and $su(5)$ cases. Note that this formula, that is
valid for all representation Hilbert spaces, is characterized
precisely by a right number (algebra rank) of the representation
labels. As a first and most precious application, based on this
generalized Gell-Mann formula, we obtained for the first time a
closed form of the generic expressions of all matrix elements of
the $sl(5,\R)$ noncompact generators. A distinct feature of our
generalized Gell-Mann formula approach is that the resulting
expression goes beyond the standard notion of a deformation of the
contracted algebra, as it depends on additional operators not
belonging, however directly related, to the contracted algebra.
Due to this fact, our generalization of the Gell-Mann formula is
remarkable simple (compared to complicated polynomial expressions
appearing in some other approaches to generalize the Gell-Mann
formula), nevertheless establishing a direct relation between
representations of the contracted and original algebras.

\section{Acknowledgments}

This work was supported in part by MNTR, Project-141036. One of
us, I.S., would like to acknowledge hospitality and useful
discussions at the Institute for Nuclear Research and Nuclear
Energy in Sofia (Bulgaria) during his visit as early stage
researcher supported by the FP6 Marie Curie Research Training
Network "Forces-Universe" MRTN-CT-2004-005104.

\end{document}